# Internet and political communication – Macedonian case

MSc. Sali Emruli[1], Prof. dr. sc. Miroslav Bača [2]

**Faculty of Organization and Informatics, University of Zagreb,**
**Varaždin, 42000, Croatia**

**Faculty of Organization and Informatics, Department for biometrics, University of Zagreb,**
**Varaždin, 42000, Croatia**

## Abstract

Analysis how to use Internet influence to the process of political communication, marketing and the management of public relations, what kind of online communication methods are used by political parties, and to assess satisfaction, means of communication and the services they provide to their party's voters (people) and other interest groups and whether social networks can affect the political and economic changes in the state, and the political power of one party.

## 2. Introduction

The analysis will be done in a way that will create a list of largest political parties in the Republic of Macedonia, their communications infrastructure through ICT (website), content and the manner in which they placed their information and receive feedback from voters.

Internet, social networking, Web 2.0, Facebook, YouTube, blog ... All these are relatively new word in the political vocabulary, new concepts, new media and new opportunities for the transmission of ideas and messages are not enough channels used to communicate with the public. Although the practice of using the Internet in local political advertising goes back to the nineties, only in recent years the advent of new tools and social networks demonstrates true strength of this medium.

Besides direct access to the public, political ideas, it provides full force confrontation, but also provides a relatively convenient ground for review of public attitudes, research and development of certain ideas. Using such a change in social communication, transmission of political messages through the transition from traditional forms of communication and finding new paths to the recipients. Professional and political public for years following the development of the Internet as a medium, but he showed the greatest strength in the last U.S. presidential election.

Political power depends on the satisfaction of the people towards a particular party and party connections with other parties or organizations. Well-developed social network provides further prestige and power of the party and its direct channel of communication with voters and other influential interests groups.

## 3. Political Communication through Internet

Internet and politics in the modern world have become inseparable and thus gradually eliminating barriers to free flow of information between the political decision-makers and those in whose name the benefits they bring (the public). Countries in transition must follow the contemporary trends of fitting of the Internet in the area of political communication, which simultaneously causes the change to the model that is still the dominant, of political communication based on secrecy and lack of transparency.

In Macedonia, the network and politics are still not together, except in the case of international organizations. Internet is not fully incorporated into political communication (or, more precisely, it is not done properly). A key condition (requirement) for this is application of technology and simultaneous transformation of consciousness. This change requires the rejection of the principle of confidentiality as a condition of political activity of government and party, because it is absolutely contrary to the nature of the Internet. It is necessary also to strengthen the awareness of the importance of on-line crystallization of public opinion, and more intensive and better connection of on-line and off-line political stage.

## 4. Impact of Social network analysis in politics

Political communication is a new and exciting area of research and teaching that is located at the crossroads of the study of communication, political parties and electoral behavior. As well as profiling the changing nature of the media system such an approach invariably leads us onto what we term the 'new political communication' - that based around the new Information and Communication Technologies (ICTs). We examine the work that has been done on the uses of the new media by parties and politicians across a range of democratic contexts and offer some insights into the strong challenges they introduce for the established manufacturers of political communication.
One of the key uses of the Internet is to build databases of voter data and access that through different applications for different purposes. Because data entry can be easily done automatically by scanners or by hand more campaigns and political operatives are recognizing the importance of capturing, storing, analyzing and using voter information. What used to take days of analyzing can now take minutes by using computers to analyze important information. That data can also be used offline or online for a number of different ways and the usage of these systems have become key components of the political system.

Throughout history political campaigns have evolved around the advancing technologies that are available to candidates. As technology develops, candidates are able to permeate the lives of citizens on a daily basis. Television, radio, newspapers, magazines, billboards, yard signs, bumper stickers, and Internet websites all create a means of spreading political platforms.

While the traditional forms of media are still an integral portion of campaign strategy, the availability of the Internet opens the door of campaign tools waiting for candidate's attention. The Internet provides numerous opportunities for politicians to reach the polity. Among those is a new phenomenon called social networking websites. Social networking sites have gained popularity in the last few years. These sites are growing popular particularly on college campuses nationwide. Specifically social networking websites such as MySpace and Facebook have provided users with a new form of communication. When new forms of communication are made available, political candidates begin to use the new technology to their advantage. What social networking websites allow politicians to do is to create a sense of personalized communication with their constituents. This personalization of politics enables voters and politicians alike to feel as though a connection is made. The Internet can make direct communication possible among government officials, candidates, parties, and citizens. As history shows us, when new technologies are made available, they begin to reshape the personalization factor between the candidate and the voter. This increase in interpersonal interactivity has shown to offer opportunities and increase success for political campaigns.

## 5. Political Parties in Republic of Macedonia
## 5.1. Overview of the political system

Macedonia is a Republic having multi-party parliamentary democracy and a political system with strict division into legislative, executive and judicial branches. From 1945 Macedonia had been a sovereign Republic within Federal Yugoslavia and on September 8, 1991, following the referendum of its citizens, Macedonia was proclaimed a sovereign and independent state. The Constitution of the Republic of Macedonia was adopted on November 17, 1991, by the first multiparty parliament. The basic intention was to constitute Macedonia as a sovereign and independent, civil and democratic state and also to create an institutional framework for the development of parliamentary democracy, guaranteeing human rights, civil liberties and national equality.

The Assembly is the central and most important institution of state authority. According to the Constitution it is a representative body of the citizens and the legislative power of the Republic is vested in it. The Assembly is composed of 120 seats.

The President of the Republic of Macedonia represents the Republic, and is Commander-in-Chief of the Armed Forces of Macedonia. He is elected in general and direct elections, for a term of five years, and two terms at most.

Executive power of the Republic of Macedonia is bicephalous and is divided between the Government and the President of the Republic. The Government is elected by the Assembly of the Republic of Macedonia by a majority vote of the total number of Representatives, and is accountable for its work to the Assembly. The organization and work of the Government is defined by a law on the Government.

In accordance with its constitutional competencies, executive power is vested in the Government of the Republic of Macedonia. It is the highest institution of the state administration and has, among others, the following responsibilities: it proposes laws, the budget of the Republic and other regulations passed by the Assembly, it determines the policies of execution of laws and other regulations of the Assembly and is responsible for their execution, decides on the recognition of states and governments, establishes diplomatic and consular relations with other states, proposes the Public Prosecutor, proposes the appointment of ambassadors and representatives of the Republic of Macedonia abroad and appoints chiefs of consular offices, and also performs other duties stipulated by the Constitution and law.

In Macedonia there are more political parties participating in the electoral process at national and local level.

## 5.2. Current Structure

Parties of traditional left and right:
Coalition VMRO – DPMNE (63 mandates, right oriented Macedonian party)
Democratic Party of the Albanians (12 mandates, right oriented Albanian party)
Coalition "SONCE" – SDSM (27 mandates left oriented Macedonian party)
Democratic Union for Integration (18 mandates left oriented Albanian party)

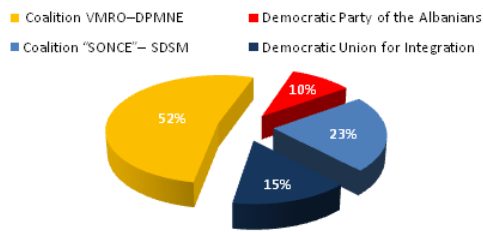

Fig. 1 Current political structure in Republic of Macedonia and Political Parties mandate percentage win in latest parliamentary elections.

## 6. Political party web sites and use of social media

Party websites represent an application of a technology which has led those dealing in votes to invest considerable amounts of time and money. This part presents a survey of the websites of Macedonian political parties. It examines the individual parties on the web and the party system on the web as virtual counterparts of the ordinary parties and party system.

Because of the importance of applying information and communication technologies (ICT) in the work of central and local government and in terms of facilitating the life and work of citizens, as well as the major role in the development of information society in EU integration process, an analysis has been conducted on websites and social media usage by political parties in Macedonia.

The research includes analysis of websites, analysis of the use of social media and online activities compared in terms of seats obtained by political parties.

In account were taken only websites of the major political parties that can certainly be determined to be guided by their info centers.

In order to level the differences between the parties and a common base for comparison, two indexes were created: Internal platform which are the party's official website and External platforms which are websites that defined as social media platforms.

While reviewing the websites I have searched for features that can be considered as social media features and that were incorporated in parties' websites.

Contrary to traditional content analysis where texts are the subject of a thorough analysis, here a content analysis was made, but on a more general level of website sections and less on their content.

Platforms that are not owned by the parties and considered as social media platforms. On these platforms the party has an official profile/user that is uploading the content and has the permissions to monitor moderate the other users' activity.

The results had showed that the relatively popular parties, The VMRO-DMPNE, Social Democrats, DPA, DUI and ND had been using more social media features and platforms than the other parliament parties.

Despite these findings there are no clear signs of an established use of social media as a political communication strategy. There is not enough correlation between number of parliament members and website's users. Some of major parties are the worst in translating voters into website's users.

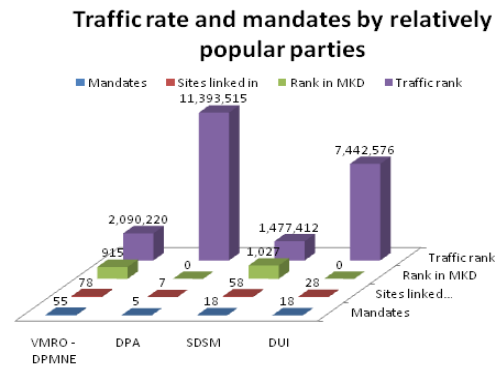

Fig. 2 Represents Traffic rate and mandates by relatively popular parties

## 6.1. Purpose of research

Due to the increased use of Internet in Macedonia and the increased influence of the same as medium, research has been conducted to determine how political parties use the influence of Internet.

The study aims to evaluate several levels of Internet usage and social networks in everyday political events:

- To determine the level of quality and implementation of the standards of the political parties websites
- To determine the level of use of social media for their promotion
- To determine the level of online communication of political parties with public
- To compare the number of online supporters with conquered mandate of the last elections.

One area of running a political campaign is Internet, and also the use of social media. The analysis will include internal platform which are the party's official website their social content and does this web pages have links to the external platforms which are websites that defined as social media platforms. Also, the technical characteristics of the sites.

### 6.1.1. Research questions and hypotheses

The analysis of political parties' websites is based on clearly defined issues that are divided into several categories, in order to evaluate every aspect of the content and making the website of the political party. Questionnaire for this section is designed to address the following main questions:

- In what languages website is available
- What type of content is offered on website (text, multimedia, transparency ...)
- Applying the standard for usability of website
- Usage of Social Media

### 6.1.2. Limits

In conducting the research and when creating the list of political parties whose Web sites will be analyzed, there were taken in consideration only those whose identity could be confirmed. It means that it is evident that website is managed by the political party.

### 6.1.3. Methodology

To implement this part of the research content analysis methodology was used. Drawing up the list of web sites that will be use, their analysis is processed by strictly defined form, with concrete questions and directions.

The form consists of three main issues, which contain additional questions about obtaining the necessary information and conclusions.

The main issues are:

- In which language versions websites are accessible
- What type of content is offered by websites
- What kind of social media are used by political party

### 6.2. Language version

The websites analysis will include Web sites of relatively popular political parties in Macedonia. From 19 political parties, only 13 (70%) have their own web sites for promotion and marketing of their political activity. From analyzed Web sites, only 4 offer bilingual accessibility (30%), and others offer information only in the language of own ethnicity (unilingual content).

| Language versions of political party websites | Political party | Percentage |
|---|---|---|
| Macedonian | 10 | 77% |
| Albanian | 3 | 23% |
| English | 4 | 30% |
| Turkish | 1 | 8% |
| Serbian | 1 | 8% |

Table 1: Language Versions

Fig. 3 Language versions

As I mentioned above from analyzed websites, only 4 offer bilingual accessibility (30%), and nine other political parties (70%) offer only unilingual content.

| Multilanguage Usage by political party websites | Political party | Percentage |
|---|---|---|
| Multilanguage | 4 | 31% |
| Only Macedonian | 6 | 46% |
| Only Albanian | 2 | 15% |
| Only Turkish | 1 | 8% |

Table 2: Multilanguage Usage

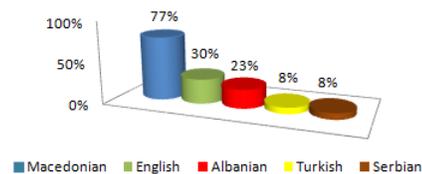

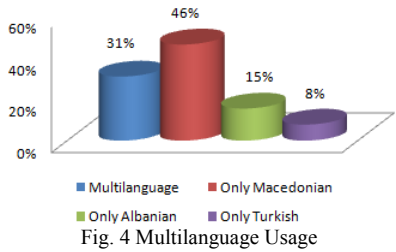

Fig. 4 Multilanguage Usage

## 6.3. Offered content

The second area for which websites analysis was performed is content offered by political party websites to their readers. The type of content is divided into text and multimedia.

| | | Mandates | Textual content | | | |
|---|---|---|---|---|---|---|
| | | | About Us | Events/Calendar | Notifications | Out Links |
| 1 | VMRO - DPMNE | 55 | 1 | 1 | 1 | 1 |
| 2 | Social - Democratic Union of Macedonia | 18 | 1 | 1 | 1 | 1 |
| 3 | Democratic Union for Integration | 18 | 0 | 1 | 1 | 1 |
| 4 | Democratic Party of the Albanians | 5 | 1 | 1 | 1 | 0 |
| 5 | Liberal - Democratic Party | 4 | 1 | 1 | 1 | 0 |
| 6 | New Democracy | 4 | 1 | 1 | 1 | 1 |
| 7 | New Social-Democratic Party | 3 | 1 | 1 | 1 | 1 |
| 8 | Socialist Party | 1 | 1 | 1 | 1 | 1 |
| 9 | Democratic reconstruction of Macedonia | 1 | 1 | 1 | 1 | 1 |
| 10 | Democratic union | 1 | 1 | 1 | 1 | 1 |
| 11 | Democratic Party of Serbs in Macedonia | 1 | 1 | 1 | 1 | 1 |
| 12 | Democratic Party of Turks in Macedonia | 1 | 1 | 1 | 1 | 0 |
| 13 | Liberal Party | 1 | 1 | 1 | 1 | 0 |

Table 3-1: Content offered by political parties websites

| | | Mandates | Multimedia | | | |
|---|---|---|---|---|---|---|
| | | | Photos | Photo gallery | Audio clips | Video Clips |
| 1 | VMRO - DPMNE | 55 | 1 | 0 | 0 | 0 |
| 2 | Social - Democratic Union of Macedonia | 18 | 1 | 1 | 1 | 1 |
| 3 | Democratic Union for Integration | 18 | 1 | 1 | 0 | 1 |
| 4 | Democratic Party of the Albanians | 5 | 1 | 0 | 0 | 1 |
| 5 | Liberal - Democratic Party | 4 | 1 | 1 | 0 | 1 |
| 6 | New Democracy | 4 | 1 | 1 | 1 | 1 |
| 7 | New Social-Democratic Party | 3 | 1 | 1 | 0 | 1 |
| 8 | Socialist Party | 1 | 1 | 0 | 0 | 0 |
| 9 | Democratic reconstruction of Macedonia | 1 | 1 | 1 | 0 | 1 |
| 10 | Democratic union | 1 | 1 | 0 | 0 | 0 |
| 11 | Democratic Party of Serbs in Macedonia | 1 | 1 | 1 | 1 | 1 |
| 12 | Democratic Party of Turks in Macedonia | 1 | 1 | 0 | 0 | 0 |
| 13 | Liberal Party | 1 | 1 | 0 | 0 | 0 |

Table 3-2: Content offered by political parties' websites

The results of this part of the research shows that most of the websites of political parties are filled with textual content, but that the textual content is not linked to the outside source (Out Links) 31% of websites, while regarding the multimedia content nearly 46% of the websites of political parties have no photo gallery, 77% of websites of political parties have no audio clips and 38% of websites have no video clips. Also a lack in all the websites of political parties is informing the guests for future activities and events. Most political parties use their websites to archive articles from the media, rather than used to inform their supporters.

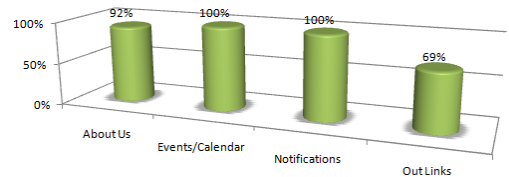

Fig.5 Textual content of all political party web sites

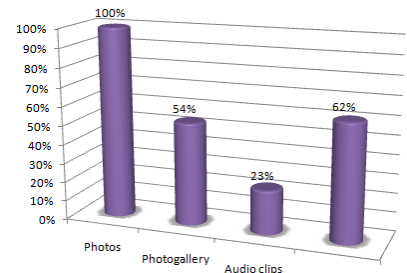

Fig. 6 Multimedia content of all political party web sites

Frequently political parties put banners to external WebPages, and not applying the concept of linking in the text. Although most of the texts offered by political parties at their web sites are excerpts from articles in the media, although they cite the source from where the content is downloaded, they not publish the link to the original article, not even the online edition of the medium.

Besides textual content almost all political parties are offering and multimedia content. Most of political parties have placed videos of nearly any report or television interview. Besides the video clips, several political parties offer galleries of their activities. Only audio clips are missing from multimedia content on political party websites. Only three political parties have offered this type of content, and they have offered several songs (hymns of the party) for download, but they were not taken into account.

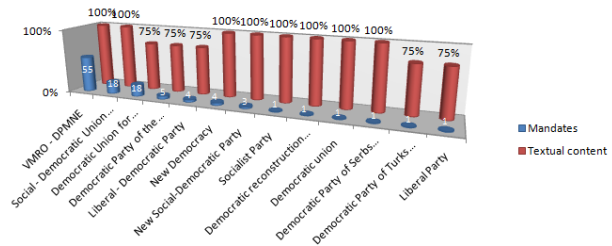

Fig.7 Textual content of each political party websites

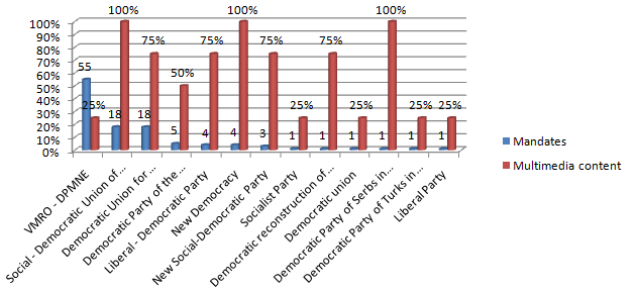

Fig.8 Multimedia content of each political party website.

Besides the types of content (text and multimedia) in this part of the analysis were explored also the ways of communication which are offered by political parties through their websites.

Surprising fact that political parties do not use their sites for the opportunity to communicate and besides the possibility of sending e-mails almost there is no other possible way to contact them, except a very small number of political parties which offers email and other contact information as phone number, address and the like.

Very few political parties have set forum on their website, also very small number of political parties offers an opportunity for asking questions and publishing answers online.

| | Mandates | Public e-mail | Form | Phone | ZIP address | Mailing list | Poll | Discussion | Forum |
|---|---|---|---|---|---|---|---|---|---|
| VMRO - DPMNE | 55 | 0 | 1 | 1 | 1 | 0 | 0 | 1 | 1 |
| Social - Democratic Union of Macedonia | 18 | 1 | 1 | 1 | 1 | 1 | 1 | 1 | 1 |
| Democratic Union for Integration | 18 | 1 | 1 | 1 | 1 | 1 | 0 | 0 | 0 |
| Democratic Party of the Albanians | 5 | 1 | 1 | 1 | 0 | 1 | 0 | 0 | 0 |
| Liberal - Democratic Party | 4 | 1 | 1 | 1 | 1 | 0 | 0 | 0 | 0 |
| New Democracy | 4 | 1 | 1 | 0 | 0 | 0 | 0 | 1 | 1 |
| New Social-Democratic Party | 3 | 1 | 1 | 1 | 1 | 1 | 0 | 0 | 0 |
| Socialist Party | 1 | 1 | 0 | 1 | 1 | 0 | 0 | 0 | 0 |
| Democratic reconstruction of Macedonia | 1 | 1 | 1 | 1 | 1 | 1 | 0 | 0 | 0 |
| Democratic union | 1 | 1 | 1 | 0 | 0 | 0 | 0 | 0 | 0 |
| Democratic Party of Serbs in Macedonia | 1 | 1 | 1 | 1 | 0 | 0 | 1 | 0 | 0 |
| Democratic Party of Turks in Macedonia | 1 | 1 | 1 | 1 | 0 | 0 | 0 | 0 | 0 |
| Liberal Party | 1 | 1 | 1 | 1 | 1 | 1 | 0 | 0 | 0 |

Table 4: Ways of communication offered by political parties

Table shows that in terms of interaction, political parties are not handled and did not use the opportunities of new media field. Besides basic information such as postal address, phone and email address, no other method is used. Sites of some political parties have disabled the opportunity to contact them via e-mail or form, but they offer only the traditional ways of communication (telephone and letter).

For transparency of the website is necessary to enable seeing the number of visitors on the site, which was also left out of more websites of the political parties. A very small part of the political parties had included counters on their websites, whether public or just used by the administrators of the website. This means that political parties do not take care of attendance (visitors) of their websites.

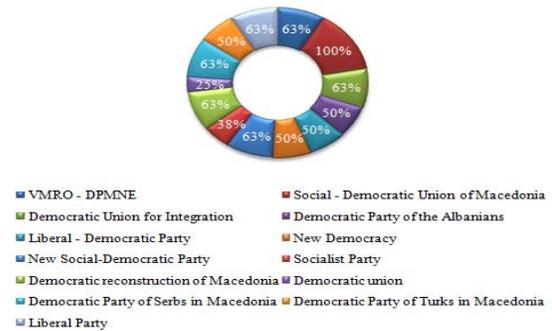

Fig.9 Opportunities for contact / interaction / transparency offered by each political party in Macedonia

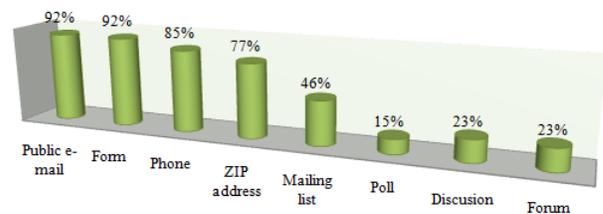

Fig.10 Opportunities for contact / interaction / transparency offered by all political party in Macedonia by category

| | Mandates | Visitors Counter |
|---|---|---|
| VMRO - DPMNE | 55 | 0 |
| Social - Democratic Union of Macedonia | 18 | 0 |
| Democratic Union for Integration | 18 | 0 |
| Democratic Party of the Albanians | 5 | 0 |
| Liberal - Democratic Party | 4 | 0 |
| New Democracy | 4 | 0 |
| New Social-Democratic Party | 3 | 0 |
| Socialist Party | 1 | 0 |
| Democratic reconstruction of Macedonia | 1 | 0 |
| Democratic union | 1 | 0 |
| Democratic Party of Serbs in Macedonia | 1 | 1 |
| Democratic Party of Turks in Macedonia | 1 | 0 |
| Liberal Party | 1 | 0 |
| Total | | 1 |
| Percentage | | 8% |

Table 5: Visitors Counter on political web pages

From the table we can conclude that almost 92% of political parties have no counters on their websites, as a consequence of lack of counters we cannot say with certainty about the attendance (visitors) of website of certain political parties.

Almost all political parties have used CMS (Content Management System) for making their websites, so they meet the basic rules for usability of the website. However astonishing fact that despite meeting the technical specifications for usability, they have errors that are not inherent for the platforms that are used, for example the search box which does not work properly and the like. In terms of recommendations for visibility of search engines, many political parties does not satisfy the conditions, which means that their site search will not be among the first results and will not be easily accessible to readers.

## 6.4. Technical Specifications of political parties' websites

This part of research covers the technical characteristics of the websites of political parties, respectively hosting and platform on which websites are set up and registration of the domain are shown in the table below.

| | Mandates | Software solution | Platform | Server | Registered in | Hosted in |
|---|---|---|---|---|---|---|
| VMRO - DPMNE | 55 | ASP | Windows 2003 | Microsoft-IIS/6.0 | MK | MK |
| Social - Democratic Union of Macedonia | 18 | ASPX | Windows 2000 | Microsoft-IIS/5.0 | MK | MK |
| Democratic Union for Integration | 18 | PHP | Windows 2003 | Microsoft-IIS/6.0 | MK | MK |
| Democratic Party of the Albanians | 5 | PHP | Unknown | Apache/2.2.X OVH | France | France |
| Liberal - Democratic Party | 4 | ASP | Windows 2000 | Microsoft-IIS/5.0 | MK | MK |
| New Democracy | 4 | ASP | Windows 2003 | Microsoft-IIS/6.0 | US | US |
| New Social-Democratic Party | 3 | PHP | Unknown | Unknown | MK | MK |
| Socialist Party | 1 | PHP | Unknown | Unknown | MK | NL |
| Democratic reconstruction of Macedonia | 1 | PHP | Unknown | Unknown | MK | US |
| Democratic union | 1 | PHP | Windows 2000 | Microsoft-IIS/5.0 | MK | MK |
| Democratic Party of Serbs in Macedonia | 1 | PHP | Unknown | Unknown | US | US |
| Democratic Party of Turks in Macedonia | 1 | PHP | Unknown | Unknown | US | US |
| Liberal Party | 1 | PHP | Unknown | Unknown | US | US |

Table 6: Technical Specifications of political parties web sites

If we do comparison of software solutions which is more used we can conclude that 69% of political parties websites use the PHP programming language, 23% use ASP programming language and only 8% use ASPX programming language. Platforms that are used by political parties websites are 23% Windows Server 2003, 23% of them use Windows 2000 platform and 54% were Unknown platforms.

| Software solution | | Platform | |
|---|---|---|---|
| ASPX | 8% | Windows 2003 | 23% |
| ASP | 23% | Windows 2000 | 23% |
| PHP | 69% | Unknown | 54% |

Table 7: Software solutions and platforms used by political parties websites

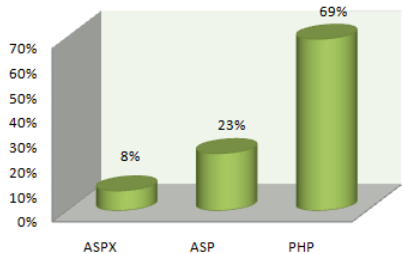

Fig.11 Software solution used by political parties' websites in Macedonia

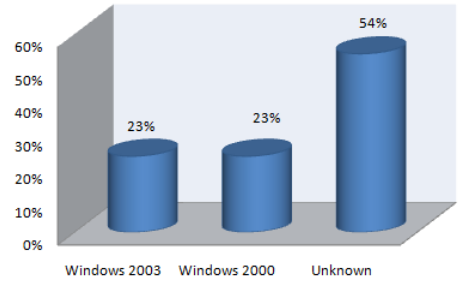

Fig.12 Platforms used by political parties' websites in Macedonia

| Server | | Hosted In | |
|---|---|---|---|
| Microsoft-IIS/6.0 | 23% | Macedonia | 46% |
| Microsoft-IIS/5.0 | 23% | US | 38% |
| Apache/2.2.X OVH | 8% | France | 8% |
| Unknown | 46% | NL | 8% |

Table 8: Servers used by political parties' websites in Macedonia

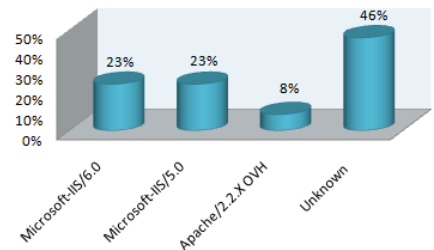

Fig.13 Servers used by political parties' websites in Macedonia

Most of political parties websites are hosted in Macedonia 46% of the total number of political Web sites, 38% of them are hosted in the US, 8% are hosted in France and 8% are hosted in the Netherland.

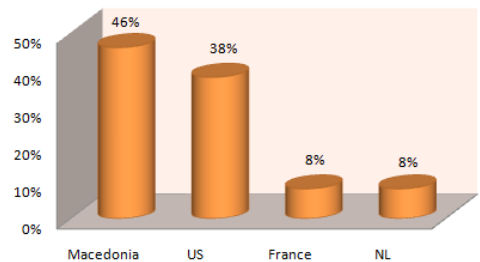

Fig.14 Hosted in Countries

Most of the websites of political parties are registered in the Republic of Macedonia 63%, while 31% are registered in the United States and 7% are registered in France.

| Registered In | |
|---|---|
| Macedonia | 62% |
| US | 31% |
| France | 7% |

Table 9: Registered in countries

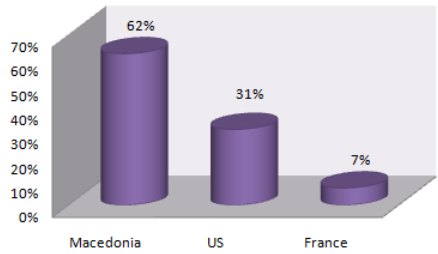
Fig.15 Registered in countries

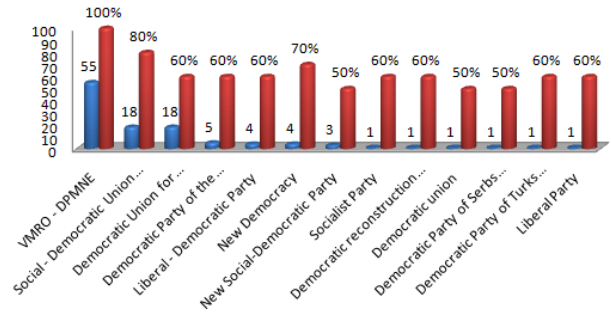
Fig.16 Usage of social media by each political party

According the table in which the technical characteristics of the websites of political parties are presented, we can see that political parties have used free solutions. While the information for domains show that websites of political parties often are registered and hosted in Macedonia, but there are a few exceptions, some political parties have their sites hosted in the U.S., but again Macedonia is dominant in terms of domain registration. Here it is important to mention that those domains that are not registered in Macedonia have no extension. mk.

## 6.5. Profiles on social media websites

Recently, many popular are accounts of political parties and politicians on social media, so precisely because they were included in this study. According to this fact i have collected information which politicians (party presidents) have their official account on social media (social networking sites) and the political parties use these social media to communicate with electors.

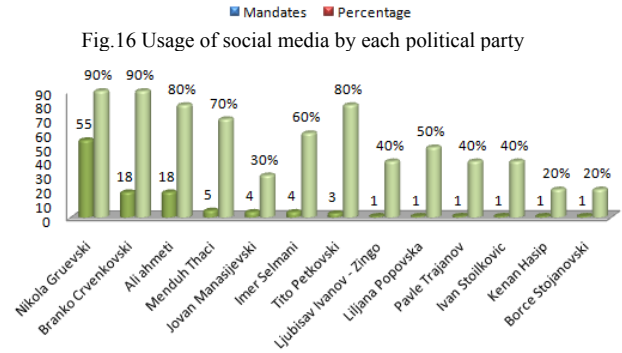
Fig.17 Usage of social media by each president of political parties

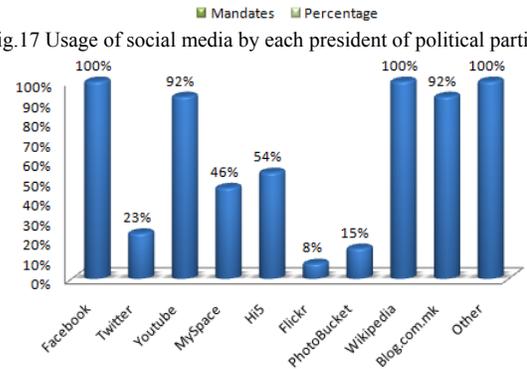
Fig.18 Usage of social media by all political parties

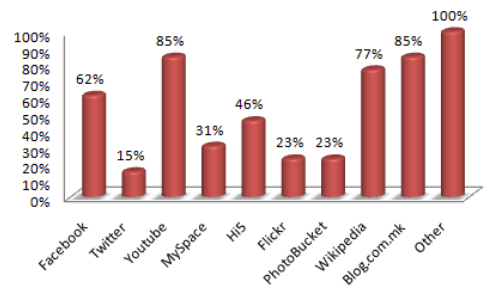
Fig.19 Usage of social media by all presidents of political parties

Table 10: Usage of social media by each political party and its leader

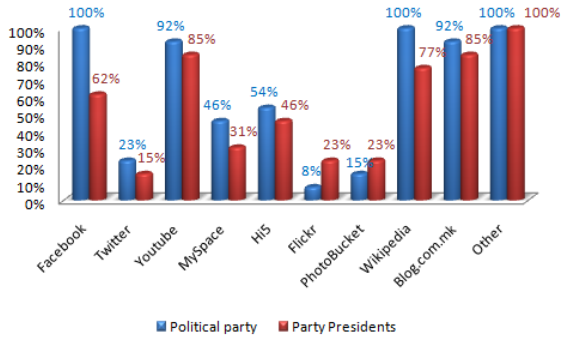

Fig.20 Usage of social media by all political parties VS all party presidents

According to the results, they show that despite the popularity of social media (SNS), political parties and politicians (presidents of the parties) have not used fully the advantages and opportunities of these tools to connect with their supporters. Most popular services are Facebook and YouTube, while platforms for sharing images were used rarely. Some political parties had implemented photo galleries on their sites.

## 7. Conclusion

The Internet first became a significant political tool in offering one-way communication for political parties with the development of political parties' websites. However, politics online is no longer as valued for its one-way communication but is now praised, and used for the opportunities it provides to conduct two way communications between political parties, their campaigns, and potential voters.

In today's political arena, websites and Internet resources, such as weblogs, social networks, podcasts and compatible video formats are being shared as a means of consuming and disseminating information via the web. As a result, websites are becoming a major if not the number one resource for political campaigns to contact supporters, volunteers, and donations. At the same time, for the consumer, or, in this case, the voter, the Internet has become a primary resource for campaign media explored via web blogging, campaign websites, news sites, social networks, video sharing and podcasts. While traditional websites are still offering significant value to the political world, technology is pushing the envelope steps further with the use of web blogging, the development of social networks, the availability of podcasts (news and opinion related), and video sharing through sites such as YouTube, which provide the general public with video clips (of up to 5+ minutes in length). Thus the issue is no longer whether politics is online but, instead, in what form and with what consequences.

Politics on the Internet has expanded beyond static two-dimensional web pages that used to serve as online billboards, flyers for a candidates position, and the traditional barriers of physical organizing. This has ushered in a new era of online consumer media and networking content that is saturated by political and campaign content. Furthermore, the phenomena of campaigns and the Internet is becoming less about what is featured on the campaign website, and instead consists more of user generated and user spread content that circulates virally on the Internet, connecting supporters from across the globe.

So far, Web 2.0 has had a weak e-ruptive effect on Macedonian party politics. On the horizontal dimension, the parties' share of activities on Web 2.0 has mainly followed what could be expected from their share of votes in 2008 parliamentary election. But deviations from the pattern indicate there are variable priorities, meaning that given a minimum of resource, parties and activists can decide to be "big in Web 2.0 politics" or decide not to. Regarding the vertical dimension of e-ruption, it appears that the national party organizations has gained more control and initiative in 2009 presidential an municipality election, the more anarchical situation of 2008 being temporary, due to sudden introduction of new technologies. Furthermore, while the number of users, viewers, members, followers and bloggers may have doubled since 2008, the party political Web 2.0 segment is still very small. This is both as a segment on Web 2.0 and as segment of voters in general.

Therefore the Web 1.5 hypothesis appears to give the best description. Furthermore, a likely next step is an even more integrated and proactive strategy, as indicated by providing guidance and cues on the party web sites, as well as setting up party specific networks or "zones" on places like Facebook. Success stories of internet politics, and especially Obama, have had a significant impact on Macedonian media. Comments like the one quoted below is quite common:
"Macedonian politicians have a lot to learn from Obama and his staff when it comes to running electoral campaigns. In particular, they should notice his priority of digital media, a part of the campaign which can be run without especially high costs."

Party strategists have also been inspired by the American experience. However, to get Macedonian voters drawn into Web politics in sufficient numbers in the first place, a more systemic approach is called for. During the American presidential campaign common "entrances" or "portals" to party politics on the main Web 2.0 sites were set up on established sites as on Facebook, YouTube and Twitter.

Some differences between the American and Macedonian party systems should also be noted. A national party in the USA and Macedonia is quite simply different entities. Population- and territorial size, as well as diversity, place different demands on local networking and autonomy, as well as effective coordination and communication between the localities. American parties also have a much looser structure, with relatively few members and dormant local branches. Macedonian parties on the other hand are still are relatively strong organizations and less reliant on ad hoc networking. Thirdly, American elections are candidate-centered, in contrast to the party centered approach found in Macedonia. These differences may be reduced over time, as Macedonia – along with other European countries – is approaching a model with decoupled local branches, fewer members and more focus on individual leaders. But they are still significant enough to warrant the question whether Web 2.0 is more functional for American parties and therefore more "rational" to use for winning elections, exactly because these parties are more like network parties in the first place.

As such, it may therefore seem like a paradox that it is the SDSM and VMRO-DPMNE which have most fully embraced Web 2.0. They are one of the oldest parties and probably still have the most effective and vital party organization. However, this also means that the party has the resources and structure to effectively implement their Web 2.0 presence, provided the party leadership thinks it necessary. It is another useful media channel for communicating with members and voters.

The Internet is a unique forum for politics as it provides back and forth communication and allows for an exchange of information between users and sources. The Internet also offers its users greater access to information and the ability to express themselves in various online political arenas. In addition, individuals use the Internet as a tool to find and join groups that share their similar ideological, cultural, political and lifestyle preferences.

**S. Emruli,** received his bachelor degree from Faculty of Communication Sciences and Technologies in Tetovo SEE University (2006), MSc degree from Faculty of Organization and Informatics, Varaždin (2010). Currently works as professional IPA Advisor at Ministry of Local Self Government in Macedonia.

**M. Bača**, is currently an Associated professor, University of Zagreb, Faculty of Organization and Informatics. He is a member of various professional societies and program committee members, and he is reviewer of several international journals and conferences. He is also the head of the Biometrics centre in Varaždin, Croatia. He is author or co-author more than 70 scientific and professional papers and two books.